\documentclass[journal]{IEEEtran} 				
\IEEEoverridecommandlockouts 						
\usepackage{amsmath,amssymb}
\usepackage{amsthm}
\usepackage{color}
\usepackage[mathscr]{eucal}
\usepackage{graphics,graphicx,multicol}
\usepackage{epsfig}
\usepackage{enumerate}
\usepackage{subfigure}
\usepackage{algorithmic}
\usepackage{algorithm}
\usepackage{morefloats}
\usepackage{enumerate}
\usepackage{subfigure}
\usepackage{multirow}
\usepackage{xfrac}
\usepackage{array}
\usepackage{pstricks, pst-node, pst-plot, pst-circ}
\usepackage{pst-3dplot,pstricks-add,pst-solides3d,pst-grad}
\usepackage{moredefs}
\usepackage{courier}
\usepackage{cite}

\begin{document}

\title{An Optimal Resource Allocation with Frequency Reuse in Cellular Networks}
\author{Ahmed Abdelhadi and T. Charles Clancy \\
Hume Center, Virginia Tech, Arlington, VA, 22203, USA\\
\{aabdelhadi, tcc\}@vt.edu
}
\maketitle

\begin{abstract}
In this paper, we introduce a novel approach for optimal resource allocation with frequency reuse for users with elastic and inelastic traffic in cellular networks. In our model, we represent users' applications running on different user equipments (UE)s by logarithmic and sigmoid utility functions. We applied utility proportional fairness allocation policy, i.e. the resources are allocated among users with fairness in utility percentage of the application running on each mobile station. Our objective is to allocate the cellular system resources to mobile users optimally from a multi-cell network. In our model, a minimum quality-of-service (QoS) is guaranteed to every user subscribing for the mobile service with priority given to users with real-time applications. We show that the novel resource allocation optimization problem with frequency reuse is convex and therefore the optimal solution is tractable. We present a distributed algorithm to allocate the resources optimally from Mobility Management 
Entity (MME) to base stations (BS)s sectors. Finally, we present the simulation results for the performance of our rate allocation algorithm.
\end{abstract}

\begin{keywords}
Optimal Resource Allocation, Frequency Reuse, Utility Proportional Fairness
\end{keywords}

\providelength{\AxesLineWidth}       \setlength{\AxesLineWidth}{0.5pt}%
\providelength{\plotwidth}           \setlength{\plotwidth}{8cm}
\providelength{\LineWidth}           \setlength{\LineWidth}{0.7pt}%
\providelength{\LineWidthTwo}        \setlength{\LineWidthTwo}{1.7pt}%
\providelength{\MarkerSize}          \setlength{\MarkerSize}{3.5pt}%
\newrgbcolor{GridColor}{0.8 0.8 0.8}%
\newrgbcolor{GridColor2}{0.5 0.5 0.5}%

\section{Introduction}\label{sec:intro}

In recent years, the cellular system is witnessing a massive increase in real-time applications. This increase needs to be modeled mathematically for providing efficient resource allocation and scheduling methods. The problem of resource allocation for cellular systems has been under investigation for the last two decade. However, most of the research work conducted focuses on concave utility functions for achieving global optimal solutions. Otherwise, the achieved solution is not optimal. The concave utility functions lack to describe real-time applications as shown in \cite{DBLP:conf/pimrc/TychogiorgosGL11}. But the sigmoid utility function provides a very good representation of these applications. However, most of the work conducted on sigmoid utility functions provide sub-optimal or approximate solutions for resource allocation optimization problem, e.g.  \cite{DBLP:conf/globecom/TychogiorgosGL11, DL_PowerAllocation} provides approximate allocation solution and \cite{RebeccaThesis} curve 
fits the sigmoid functions to the closest concave functions to solve with conventional methods. Others only solves for sigmoid utility function above a certain threshold to ensure concavity of the function \cite{multi_cell_RA, RA_interference}.

In addition, most of the prior work on proportional fairness resource allocation optimization considers proportional fairness over rates \cite{multi_cell_RA, opt_net_ref}. Nowadays, there is a big variety in the applications running under cellular systems operators therefore fairness should be application-based not rate-based. The best representation of application-based fairness is sigmoid utility for real-time applications and logarithmic utilities for delay-tolerant applications. In \cite{Ahmed_Utility1, Ahmed_Utility2, Ahmed_Utility3}, the authors present utility proportional fairness for a cellular system with a single carrier. The authors use logarithmic and sigmoid utility functions to represent delay-tolerant and real-time applications, respectively. In \cite{Ahmed_Utility1}, the rate allocation algorithm gives priority to real-time applications over delay-tolerant applications when allocating resources as the utility proportional fairness rate allocation policy 
is used. 

In this paper, we include both sigmoid and concave utility functions for a cellular system with multiple cells. We show how to formulate the resource allocation optimization problem with frequency reuse into a convex optimization problem. 

\subsection{Our Contributions}\label{sec:contributions}
Our contributions in this paper are summarized as:
\begin{itemize}
\item We introduce a novel rate allocation optimization problem with frequency reuse that solves for utility functions that are logarithmic and sigmoid. 
\item We show that the proposed optimization problem is convex and therefore the global optimal solution is tractable. In addition, we present a distributed rate allocation algorithm to solve it. 
\end{itemize}

The remainder of this paper is organized as follows. Section \ref{sec:Problem_formulation} presents the problem formulation. Section \ref{sec:Dual} shows optimality and duality of proposed problem. In Section \ref{sec:Algorithm}, we present our distributed rate allocation algorithm with frequency reuse for the utility proportional fairness optimization problem. Section \ref{sec:sim} discusses simulation setup and provides quantitative results along with discussion. Section \ref{sec:conclude} concludes the paper.

\section{Problem Formulation}\label{sec:Problem_formulation}

\begin{figure}[]
\centering
\includegraphics[width=1.0\linewidth]{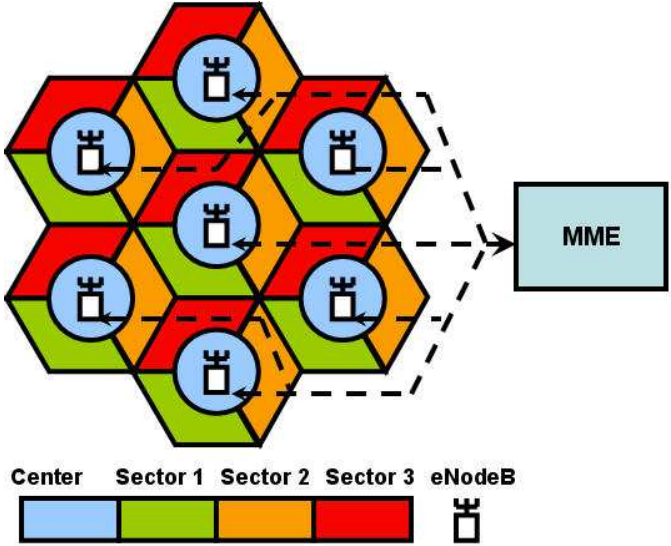}
\caption{System Model.}
\label{fig:freq_reuse}
\end{figure}

We consider a cellular network model that include cells with sectors. Therefore, we include frequency reuse. In Figure \ref{fig:freq_reuse}, we consider LTE mobile system consisting of $K$ BSs in $K$ cells each cell is divided into $L$ sector (e.g. 3 sectors) and $M$ UEs distributed in these cells. The rate allocated by the $l^{th}$ sector BS to $i^{th}$ UE is given by $r_{i}^{l}$ where $i = \{1,2, ...,M\}$, $l =\{1,2, ..., L\}$. Each UE has its own utility function $U_i(r_{i})$ that corresponds to the type of traffic being handled by the $i^{th}$ UE. Our objective is to determine the optimal rates that the $l^{th}$ carrier BS should allocate to the UEs. The utility functions are given by $U_i(r_{i})=U_i(r_{i}^{1}+r_{i}^{2}+ ...+r_{i}^{L})$ where $\sum_{l=1}^{L}r_{i}^{l} = r_i $ and $\textbf{r} =\{{r}_1, {r}_2,..., {r}_M\}$. The optimization problem for utility proportional fairness:
\begin{equation}\label{eqn:opt_prob_fairness}
\begin{aligned}
& \underset{\textbf{r}}{\text{max}} & & \prod_{i=1}^{M}U_i(r_{i}^{1}+r_{i}^{2}+ ...+r_{i}^{L}) \\
& \text{subject to} & & \sum_{l=1}^{L}r_{i}^{l} = r_i ,  \;\; \sum_{i=1}^{M}r_{i}^{1} \leq R^{1}, \;\;...\\
& & & ..., \;\; \sum_{i=1}^{M}r_{i}^{L} \leq R^{L}, \;\;  \sum_{l=1}^{L} R^{l} = R, \\
& & &  r_{i}^l \geq 0, \;\;\;\;\;l = 1,2, ...,L, i = 1,2, ...,M.
\end{aligned}
\end{equation}
where $R^{l}$ is the allocated rate by the MME to the $l^{th}$ sector, and $R$ is the sum of the allocated rates to all sectors. We assume that the same frequency band is allocated to all sectors therefore avoiding interference. So we have the assumption that $R^{l}$ is fixed for all BSs. We assume that UE can't exist in two sectors simultaneously, i.e. if $r_i^{l} \neq 0$ then $r_i^{q} = 0$ for $q \neq l$.

We assume the utility functions $U_i(r_{i}^{1}+r_{i}^{2}+ ...+r_{i}^{L})$ to be a concave or a sigmoid functions. In our model, we use the sigmoid utility function, as in \cite{DL_PowerAllocation, GAC_MILCOM}, given by
\begin{equation}\label{eqn:sigmoid}
U_i(r_{i}^{1}+r_{i}^{2}+ ...+r_{i}^{L}) = c_i\Big(\frac{1}{1+e^{-a_i(\sum_{l=1}^{L}r_{i}^l-b_i)}}-d_i\Big)
\end{equation}
where $c_i = \frac{1+e^{a_ib_i}}{e^{a_ib_i}}$ and $d_i = \frac{1}{1+e^{a_ib_i}}$. We use the logarithmic utility function, as in \cite{UtilityFairness, SAC_CA_user_discrimination}, given by
\begin{equation}\label{eqn:log}
U_i(r_{i}^{1}+r_{i}^{2}+ ...+r_{i}^{L}) = \frac{\log(1+k_i\sum_{l=1}^{L}r_{i}^l)}{\log(1+k_ir_{max})}
\end{equation}
where $r_{max}$ is the 100\% rate and $k_i$ is the rate of increase..

\section{Optimality and Dual Problem}\label{sec:Dual}
In the optimization problem (\ref{eqn:opt_prob_fairness}), since the objective function $\arg \underset{\textbf{r}} \max \prod_{i=1}^{M}U_i(r_{i}^{1}+r_{i}^{2}+ ...+r_{i}^{L})$ is equivalent to $\arg \underset{\textbf{r}} \max \sum_{i=1}^{M}\log(U_i(r_{i}^{1}+r_{i}^{2}+ ...+r_{i}^{L}))$, and given that log of sigmoid function is concave as shown in \cite{Ahmed_Utility1} then optimization problem (\ref{eqn:opt_prob_fairness}) in convex and there exists a tractable global optimal solution.

The primal problem in (\ref{eqn:opt_prob_fairness}) can be converted into the dual problem following similar steps as in \cite{Ahmed_Utility1}. The UE and BS separate optimization problems follow from similar steps as in \cite{Ahmed_Utility1, Haya_joint}. The utility proportional fairness in the objective function of the optimization problem (\ref{eqn:opt_prob_fairness}) is achieved by the MME which ensure fair allocation to the sectors to achieve equal shadow price for all users. This is done by setting $p_l=p$ for all $l$. We define the aggregated bids of the $l^{th}$ sectors by $W^l = \sum_{i=1}^{M}w_{i}^l(n)$ an so we have $W^l = p R^l$ and by summing over $l$ we have
\begin{equation}\label{eqn:MME_agg_bids}
\sum_{l=1}^{L}W^l = p \sum_{l=1}^{L} R^l = p R
\end{equation}
Then it follows from $W^l = p R^l$ and (\ref{eqn:MME_agg_bids}) that
\begin{equation}\label{eqn:MME_allc_rate}
R^l(n)=\frac{W^l}{\sum_{l=1}^{L}W^l}R
\end{equation}
which is computed in the MME and guarantees fairness.

\section{Distributed Optimization Algorithm}\label{sec:Algorithm}

The distributed resource allocation algorithm is an implementation of UE and BS dual problems, see \cite{Ahmed_Utility1, Haya_joint} for more details, and (\ref{eqn:MME_allc_rate}). Our algorithm allocates resources from multiple cells simultaneously with utility proportional fairness policy. The algorithm is divided into the $i^{th}$ UE algorithm shown in Algorithm (\ref{alg:UE_FK}) the $l^{th}$ sector algorithm shown in Algorithm (\ref{alg:BS_FK}), and MME algorithm shown in Algorithm (\ref{alg:MME_FK}). In Algorithm (\ref{alg:UE_FK}), (\ref{alg:BS_FK}) and (\ref{alg:MME_FK}), the $i^{th}$ UE starts with an initial bid $w_{i}^l(1)$ which is transmitted to the $l^{th}$ sector. The $l^{th}$ sector sends the aggregated bids from all UEs under its coverage $W^l(n) = \sum_{i=1}^{M}w_{i}^l(n)$ to MME. MME calculates the difference between the received aggregated bid $W^l(n)$ and the previously received bid $W^l(n-1)$ and exits if it is less than a pre-specified threshold $\delta$ for all sectors. 
Otherwise, MME calculates the sector rates $R^l(n)=\frac{W^l(n)}{\sum_{l=1}^{L}W^l(n)}R$ and sends it to the corresponding sectors. The $l^{th}$ sector uses $R^l(n)$ to calculate the shadow price $p_l(n) = \frac{\sum_{i=1}^{M}w_{i}^l(n)}{R^l}$ and sends its value to all the UEs in its coverage area. The $i^{th}$ UE receives the shadow prices $p_{l}$ from all sectors that covers it and calculates the new bid. This process is repeated until $|W^l(n) - W^l(n-1)|$ is less than the threshold $\delta$.

\begin{algorithm}
\caption{The $i^{th}$ UE Algorithm}\label{alg:UE_FK}
\begin{algorithmic}
\STATE {Send initial bid $w_i(1)$ to BS}
\LOOP
	\STATE {Receive shadow price $p(n)$ from BS}
	\IF {STOP from BS} %

	\STATE {Calculate allocated rate $r_i ^{\text{opt}}=\frac{w_i(n)}{p(n)}$}
			\ELSE
	\STATE {Calculate new bid $w_i (n)= p(n) r_{i}(n)$}
	\STATE {Send new bid $w_i (n)$ to BS}
		\ENDIF 
\ENDLOOP
\end{algorithmic}
\end{algorithm}

\begin{algorithm}
\caption{The $l^{th}$ sector Algorithm}\label{alg:BS_FK}
\begin{algorithmic}
\STATE {Receive sector rate $R^l(0)$ from MME}
\COMMENT{Let $R^l(0) = \frac{R}{L}$}
\LOOP
	\STATE {Receive bids $w_{i}^l(n)$ from UEs}
	\COMMENT{Let $w_{i}^l(0) = 0\:\:\forall \:i$}
		\IF {$|w_{i}^l(n) -w_{i}^l(n-1)|<\delta  \:\:\forall i$} %
	   		\STATE {Allocate rates, ${r_{i}^l}^{\text{opt}}=\frac{w_{i}^l(n)}{p_l(n)}$ to $i^{th}$ UE}  
	   		\STATE {STOP} 
		\ELSE
	\STATE {Calculate aggregated bids $W^l(n) = \sum_{i=1}^{M}w_{i}^l(n)$}
	\STATE {Send new aggregated bids $W^l(n)$ to MME}
	\STATE {Receive sector rate $R^l(n)$ from MME}
	\STATE {Calculate $p_l(n) = \frac{W^l(n)}{R^l(n)}$}
	\STATE {Send new shadow price $p_l(n)$ to all UEs}
	\ENDIF 
\ENDLOOP
\end{algorithmic}
\end{algorithm}

\begin{algorithm}
\caption{MME Algorithm}\label{alg:MME_FK}
\begin{algorithmic}
\LOOP
	\STATE {Receive bids $W^l(n)$ from $l^{th}$ sector}
	\COMMENT{Let $W^l(0) = 0\:\:\forall \:l$}
		\IF {$|W^l(n) - W^l(n-1)|<\delta  \:\:\forall l$} %
			\STATE {Allocate sectors rates, ${R^l}^{\text{opt}}=\frac{W^l(n)R}{\sum_{l=1}^{L}W^l(n)}$ to $l^{th}$ sector}  
			\STATE {STOP} 
		\ELSE
		\STATE {Calculate $R^l(n)=\frac{W^l(n)}{\sum_{l=1}^{L}W^l(n)}R$}
		\STATE {Send new sectors rates $R^l(n)$ to $l^{th}$ sector}
		\ENDIF 
\ENDLOOP
\end{algorithmic}
\end{algorithm}

\begin{figure}[]
\centering
\includegraphics[width=1.0\linewidth]{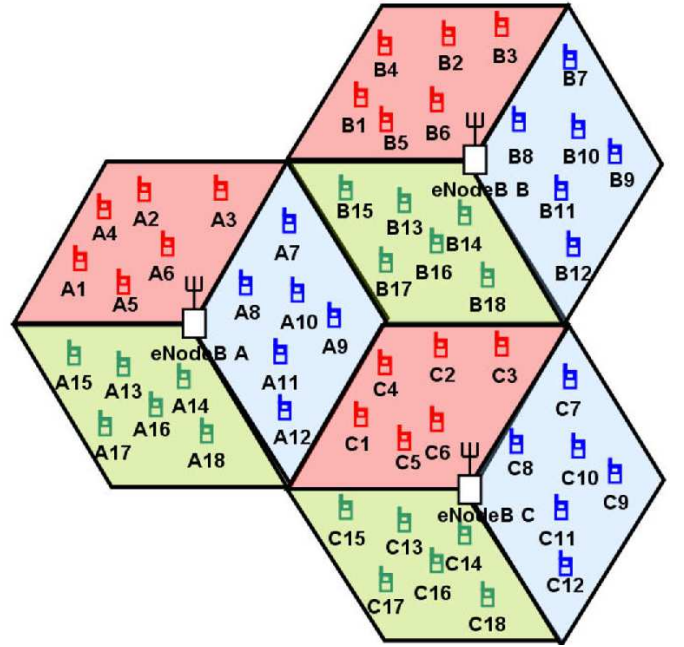}
\caption{Simulation setup with three sectors per cell.}
\label{fig:sim:System_Model}
\end{figure}
\begin{figure}[]
\centering
%
\psset{xunit=0.000909\plotwidth,yunit=0.022535\plotwidth}%
\begin{pspicture}(-61.520737,-3.888889)(1155.069124,35.818713)%


\psline[linewidth=\AxesLineWidth,linecolor=GridColor](100.000000,0.000000)(100.000000,0.532515)
\psline[linewidth=\AxesLineWidth,linecolor=GridColor](200.000000,0.000000)(200.000000,0.532515)
\psline[linewidth=\AxesLineWidth,linecolor=GridColor](300.000000,0.000000)(300.000000,0.532515)
\psline[linewidth=\AxesLineWidth,linecolor=GridColor](400.000000,0.000000)(400.000000,0.532515)
\psline[linewidth=\AxesLineWidth,linecolor=GridColor](500.000000,0.000000)(500.000000,0.532515)
\psline[linewidth=\AxesLineWidth,linecolor=GridColor](600.000000,0.000000)(600.000000,0.532515)
\psline[linewidth=\AxesLineWidth,linecolor=GridColor](700.000000,0.000000)(700.000000,0.532515)
\psline[linewidth=\AxesLineWidth,linecolor=GridColor](800.000000,0.000000)(800.000000,0.532515)
\psline[linewidth=\AxesLineWidth,linecolor=GridColor](900.000000,0.000000)(900.000000,0.532515)
\psline[linewidth=\AxesLineWidth,linecolor=GridColor](1000.000000,0.000000)(1000.000000,0.532515)
\psline[linewidth=\AxesLineWidth,linecolor=GridColor](1100.000000,0.000000)(1100.000000,0.532515)
\psline[linewidth=\AxesLineWidth,linecolor=GridColor](50.000000,0.000000)(63.200000,0.000000)
\psline[linewidth=\AxesLineWidth,linecolor=GridColor](50.000000,5.000000)(63.200000,5.000000)
\psline[linewidth=\AxesLineWidth,linecolor=GridColor](50.000000,10.000000)(63.200000,10.000000)
\psline[linewidth=\AxesLineWidth,linecolor=GridColor](50.000000,15.000000)(63.200000,15.000000)
\psline[linewidth=\AxesLineWidth,linecolor=GridColor](50.000000,20.000000)(63.200000,20.000000)
\psline[linewidth=\AxesLineWidth,linecolor=GridColor](50.000000,25.000000)(63.200000,25.000000)
\psline[linewidth=\AxesLineWidth,linecolor=GridColor](50.000000,30.000000)(63.200000,30.000000)
\psline[linewidth=\AxesLineWidth,linecolor=GridColor](50.000000,35.000000)(63.200000,35.000000)

{ \footnotesize 
\rput[t](100.000000,-0.532515){$100$}
\rput[t](200.000000,-0.532515){$200$}
\rput[t](300.000000,-0.532515){$300$}
\rput[t](400.000000,-0.532515){$400$}
\rput[t](500.000000,-0.532515){$500$}
\rput[t](600.000000,-0.532515){$600$}
\rput[t](700.000000,-0.532515){$700$}
\rput[t](800.000000,-0.532515){$800$}
\rput[t](900.000000,-0.532515){$900$}
\rput[t](1000.000000,-0.532515){$1000$}
\rput[t](1100.000000,-0.532515){$1100$}
\rput[r](36.800000,0.000000){$0$}
\rput[r](36.800000,5.000000){$5$}
\rput[r](36.800000,10.000000){$10$}
\rput[r](36.800000,15.000000){$15$}
\rput[r](36.800000,20.000000){$20$}
\rput[r](36.800000,25.000000){$25$}
\rput[r](36.800000,30.000000){$30$}
\rput[r](36.800000,35.000000){$35$}
} 

\psframe[linewidth=\AxesLineWidth,dimen=middle](50.000000,0.000000)(1150.000000,35.000000)

{ \small 
\rput[b](600.000000,-3.888889){
\begin{tabular}{c}
BS Total Rate\\
\end{tabular}
}

\rput[t]{90}(-61.520737,17.500000){
\begin{tabular}{c}
$r_i^1$\\
\end{tabular}
}
} 

\newrgbcolor{color246.004}{0  0  1}
\psline[plotstyle=line,linejoin=1,showpoints=true,dotstyle=*,dotsize=\MarkerSize,linestyle=solid,linewidth=\LineWidth,linecolor=color246.004]
(50.000000,3.893990)(100.000000,7.506888)(150.000000,9.738176)(200.000000,9.768826)(250.000000,9.819436)
(300.000000,10.219774)(350.000000,10.232818)(400.000000,10.388598)(450.000000,10.768139)(500.000000,11.037309)
(550.000000,11.218498)(600.000000,11.350100)(650.000000,11.451627)(700.000000,11.534533)(750.000000,11.606269)
(800.000000,11.667027)(850.000000,11.719334)(900.000000,11.764722)(950.000000,11.803985)(1000.000000,11.845678)
(1050.000000,11.876244)(1100.000000,11.911513)(1150.000000,11.943659)

\newrgbcolor{color247.0035}{0         0.5           0}
\psline[plotstyle=line,linejoin=1,showpoints=true,dotstyle=*,dotsize=\MarkerSize,linestyle=solid,linewidth=\LineWidth,linecolor=color247.0035]
(50.000000,0.549927)(100.000000,0.549431)(150.000000,1.763433)(200.000000,5.990559)(250.000000,9.045301)
(300.000000,10.477320)(350.000000,10.503536)(400.000000,10.795460)(450.000000,11.423878)(500.000000,11.841977)
(550.000000,12.118510)(600.000000,12.318043)(650.000000,12.471487)(700.000000,12.596557)(750.000000,12.704647)
(800.000000,12.796123)(850.000000,12.874828)(900.000000,12.943093)(950.000000,13.002129)(1000.000000,13.064799)
(1050.000000,13.110735)(1100.000000,13.163729)(1150.000000,13.212022)

\newrgbcolor{color248.0035}{1  0  0}
\psline[plotstyle=line,linejoin=1,showpoints=true,dotstyle=*,dotsize=\MarkerSize,linestyle=solid,linewidth=\LineWidth,linecolor=color248.0035]
(50.000000,0.405771)(100.000000,0.405523)(150.000000,0.664159)(200.000000,0.693006)(250.000000,0.749177)
(300.000000,3.777912)(350.000000,5.925274)(400.000000,10.090948)(450.000000,11.983152)(500.000000,12.920092)
(550.000000,13.503820)(600.000000,13.916249)(650.000000,14.230264)(700.000000,14.484770)(750.000000,14.703927)
(800.000000,14.888935)(850.000000,15.047833)(900.000000,15.185478)(950.000000,15.304395)(1000.000000,15.430532)
(1050.000000,15.522925)(1100.000000,15.629459)(1150.000000,15.726494)

\newrgbcolor{color249.0035}{0        0.75        0.75}
\psline[plotstyle=line,linejoin=1,showpoints=true,dotstyle=Bo,dotsize=\MarkerSize,linestyle=solid,linewidth=\LineWidth,linecolor=color249.0035]
(50.000000,3.893990)(100.000000,8.925258)(150.000000,11.238176)(200.000000,11.268826)(250.000000,11.319436)
(300.000000,11.719774)(350.000000,11.732818)(400.000000,11.888598)(450.000000,12.268139)(500.000000,12.537309)
(550.000000,12.718498)(600.000000,12.850100)(650.000000,12.951627)(700.000000,13.034533)(750.000000,13.106269)
(800.000000,13.167027)(850.000000,13.219334)(900.000000,13.264722)(950.000000,13.303985)(1000.000000,13.345678)
(1050.000000,13.376244)(1100.000000,13.411513)(1150.000000,13.443659)

\newrgbcolor{color250.0035}{0.75           0        0.75}
\psline[plotstyle=line,linejoin=1,showpoints=true,dotstyle=Bo,dotsize=\MarkerSize,linestyle=solid,linewidth=\LineWidth,linecolor=color250.0035]
(50.000000,0.549927)(100.000000,0.549431)(150.000000,1.763434)(200.000000,7.445426)(250.000000,10.545301)
(300.000000,11.977320)(350.000000,12.003536)(400.000000,12.295460)(450.000000,12.923878)(500.000000,13.341977)
(550.000000,13.618510)(600.000000,13.818043)(650.000000,13.971487)(700.000000,14.096557)(750.000000,14.204647)
(800.000000,14.296123)(850.000000,14.374828)(900.000000,14.443093)(950.000000,14.502129)(1000.000000,14.564799)
(1050.000000,14.610735)(1100.000000,14.663729)(1150.000000,14.712022)

\newrgbcolor{color251.0035}{0.75        0.75           0}
\psline[plotstyle=line,linejoin=1,showpoints=true,dotstyle=Bo,dotsize=\MarkerSize,linestyle=solid,linewidth=\LineWidth,linecolor=color251.0035]
(50.000000,0.405774)(100.000000,0.405526)(150.000000,0.664171)(200.000000,0.693019)(250.000000,0.749193)
(300.000000,3.801705)(350.000000,7.074362)(400.000000,11.590790)(450.000000,13.483127)(500.000000,14.420075)
(550.000000,15.003804)(600.000000,15.416235)(650.000000,15.730250)(700.000000,15.984756)(750.000000,16.203913)
(800.000000,16.388921)(850.000000,16.547820)(900.000000,16.685464)(950.000000,16.804382)(1000.000000,16.930519)
(1050.000000,17.022912)(1100.000000,17.129446)(1150.000000,17.226481)

\newrgbcolor{color252.0035}{0.25        0.25        0.25}
\psline[plotstyle=line,linejoin=1,showpoints=true,dotstyle=Bsquare,dotsize=\MarkerSize,linestyle=solid,linewidth=\LineWidth,linecolor=color252.0035]
(50.000000,3.893990)(100.000000,10.077854)(150.000000,12.738176)(200.000000,12.768826)(250.000000,12.819436)
(300.000000,13.219774)(350.000000,13.232818)(400.000000,13.388598)(450.000000,13.768139)(500.000000,14.037309)
(550.000000,14.218498)(600.000000,14.350100)(650.000000,14.451627)(700.000000,14.534533)(750.000000,14.606269)
(800.000000,14.667027)(850.000000,14.719334)(900.000000,14.764722)(950.000000,14.803985)(1000.000000,14.845678)
(1050.000000,14.876244)(1100.000000,14.911513)(1150.000000,14.943659)

\newrgbcolor{color253.0035}{0  0  1}
\psline[plotstyle=line,linejoin=1,showpoints=true,dotstyle=Bsquare,dotsize=\MarkerSize,linestyle=solid,linewidth=\LineWidth,linecolor=color253.0035]
(50.000000,0.549927)(100.000000,0.549431)(150.000000,1.763434)(200.000000,7.501706)(250.000000,12.045300)
(300.000000,13.477320)(350.000000,13.503536)(400.000000,13.795460)(450.000000,14.423878)(500.000000,14.841977)
(550.000000,15.118510)(600.000000,15.318043)(650.000000,15.471487)(700.000000,15.596557)(750.000000,15.704647)
(800.000000,15.796123)(850.000000,15.874828)(900.000000,15.943093)(950.000000,16.002129)(1000.000000,16.064799)
(1050.000000,16.110735)(1100.000000,16.163729)(1150.000000,16.212022)

\newrgbcolor{color254.0035}{0         0.5           0}
\psline[plotstyle=line,linejoin=1,showpoints=true,dotstyle=Bsquare,dotsize=\MarkerSize,linestyle=solid,linewidth=\LineWidth,linecolor=color254.0035]
(50.000000,0.405775)(100.000000,0.405527)(150.000000,0.664173)(200.000000,0.693022)(250.000000,0.749197)
(300.000000,3.807260)(350.000000,8.419442)(400.000000,13.090755)(450.000000,14.983122)(500.000000,15.920071)
(550.000000,16.503801)(600.000000,16.916232)(650.000000,17.230247)(700.000000,17.484753)(750.000000,17.703910)
(800.000000,17.888918)(850.000000,18.047817)(900.000000,18.185461)(950.000000,18.304379)(1000.000000,18.430516)
(1050.000000,18.522909)(1100.000000,18.629443)(1150.000000,18.726478)

\newrgbcolor{color255.0035}{1  0  0}
\psline[plotstyle=line,linejoin=1,linestyle=solid,linewidth=\LineWidth,linecolor=color255.0035]
(50.000000,0.291208)(100.000000,0.291079)(150.000000,0.405817)(200.000000,0.416476)(250.000000,0.436205)
(300.000000,0.737327)(350.000000,0.753415)(400.000000,0.992182)(450.000000,2.132265)(500.000000,3.828665)
(550.000000,5.739525)(600.000000,7.736054)(650.000000,9.763327)(700.000000,11.825417)(750.000000,13.973874)
(800.000000,16.109503)(850.000000,18.218780)(900.000000,20.280871)(950.000000,22.259679)(1000.000000,24.581205)
(1050.000000,26.441572)(1100.000000,28.770643)(1150.000000,31.078565)

\newrgbcolor{color256.0035}{0        0.75        0.75}
\psline[plotstyle=line,linejoin=1,linestyle=solid,linewidth=\LineWidth,linecolor=color256.0035]
(50.000000,0.288527)(100.000000,0.288400)(150.000000,0.401280)(200.000000,0.411755)(250.000000,0.431138)
(300.000000,0.726479)(350.000000,0.742241)(400.000000,0.976047)(450.000000,2.091714)(500.000000,3.752960)
(550.000000,5.626376)(600.000000,7.585655)(650.000000,9.576583)(700.000000,11.602898)(750.000000,13.715115)
(800.000000,15.815584)(850.000000,17.890849)(900.000000,19.920284)(950.000000,21.868239)(1000.000000,24.154106)
(1050.000000,25.986273)(1100.000000,28.280464)(1150.000000,30.554247)

\newrgbcolor{color257.0035}{0.75           0        0.75}
\psline[plotstyle=line,linejoin=1,linestyle=solid,linewidth=\LineWidth,linecolor=color257.0035]
(50.000000,0.285967)(100.000000,0.285841)(150.000000,0.396989)(200.000000,0.407293)(250.000000,0.426355)
(300.000000,0.716387)(350.000000,0.731851)(400.000000,0.961147)(450.000000,2.054821)(500.000000,3.684577)
(550.000000,5.524490)(600.000000,7.450440)(650.000000,9.408838)(700.000000,11.403122)(750.000000,13.482877)
(800.000000,15.551841)(850.000000,17.596621)(900.000000,19.596776)(950.000000,21.517062)(1000.000000,23.770941)
(1050.000000,25.577805)(1100.000000,27.840694)(1150.000000,30.083832)

\newrgbcolor{color258.0035}{0.75        0.75           0}
\psline[plotstyle=line,linejoin=1,linestyle=solid,linewidth=\LineWidth,linecolor=color258.0035]
(50.000000,0.283519)(100.000000,0.283395)(150.000000,0.392920)(200.000000,0.403064)(250.000000,0.421828)
(300.000000,0.706961)(350.000000,0.722151)(400.000000,0.947321)(450.000000,2.021040)(500.000000,3.622352)
(550.000000,5.432032)(600.000000,7.327902)(650.000000,9.256932)(700.000000,11.222287)(750.000000,13.272714)
(800.000000,15.313204)(850.000000,17.330424)(900.000000,19.304101)(950.000000,21.199359)(1000.000000,23.424296)
(1050.000000,25.208263)(1100.000000,27.442820)(1150.000000,29.658215)

\newrgbcolor{color259.0035}{0.25        0.25        0.25}
\psline[plotstyle=line,linejoin=1,linestyle=solid,linewidth=\LineWidth,linecolor=color259.0035]
(50.000000,0.281173)(100.000000,0.281051)(150.000000,0.389053)(200.000000,0.399048)(250.000000,0.417533)
(300.000000,0.698124)(350.000000,0.713062)(400.000000,0.934439)(450.000000,1.989934)(500.000000,3.565372)
(550.000000,5.347568)(600.000000,7.216089)(650.000000,9.118408)(700.000000,11.057441)(750.000000,13.081175)
(800.000000,15.095740)(850.000000,17.087859)(900.000000,19.037415)(950.000000,20.909868)(1000.000000,23.108428)
(1050.000000,24.871523)(1100.000000,27.080248)(1150.000000,29.270345)

\newrgbcolor{color317.0012}{0  0  1}
\psline[plotstyle=line,linejoin=1,linestyle=solid,linewidth=\LineWidth,linecolor=color317.0012]
(50.000000,0.278922)(100.000000,0.278802)(150.000000,0.385370)(200.000000,0.395225)(250.000000,0.413449)
(300.000000,0.689815)(350.000000,0.704519)(400.000000,0.922391)(450.000000,1.961150)(500.000000,3.512904)
(550.000000,5.269957)(600.000000,7.113451)(650.000000,8.991319)(700.000000,10.906250)(750.000000,12.905531)
(800.000000,14.896342)(850.000000,16.865455)(900.000000,18.792897)(950.000000,20.644438)(1000.000000,22.818807)
(1050.000000,24.562755)(1100.000000,26.747779)(1150.000000,28.914661)

\newrgbcolor{color318.0012}{0         0.5           0}
\psline[plotstyle=line,linejoin=1,linestyle=solid,linewidth=\LineWidth,linecolor=color318.0012]
(50.000000,0.276760)(100.000000,0.276641)(150.000000,0.381856)(200.000000,0.391580)(250.000000,0.409559)
(300.000000,0.681978)(350.000000,0.696465)(400.000000,0.911085)(450.000000,1.934399)(500.000000,3.464355)
(550.000000,5.198276)(600.000000,7.018741)(650.000000,8.874101)(700.000000,10.766836)(750.000000,12.743593)
(800.000000,14.712516)(850.000000,16.660425)(900.000000,18.567481)(950.000000,20.399740)(1000.000000,22.551798)
(1050.000000,24.278085)(1100.000000,26.441246)(1150.000000,28.586706)

\newrgbcolor{color319.0012}{1  0  0}
\psline[plotstyle=line,linejoin=1,linestyle=solid,linewidth=\LineWidth,linecolor=color319.0012]
(50.000000,0.274680)(100.000000,0.274562)(150.000000,0.378498)(200.000000,0.388097)(250.000000,0.405845)
(300.000000,0.674568)(350.000000,0.688852)(400.000000,0.900443)(450.000000,1.909439)(500.000000,3.419238)
(550.000000,5.131772)(600.000000,6.930940)(650.000000,8.765478)(700.000000,10.637673)(750.000000,12.593579)
(800.000000,14.542234)(850.000000,16.470504)(900.000000,18.358674)(950.000000,20.173068)(1000.000000,22.304450)
(1050.000000,24.014368)(1100.000000,26.157259)(1150.000000,28.282859)

\newrgbcolor{color320.0012}{0        0.75        0.75}
\psline[plotstyle=line,linejoin=1,linestyle=solid,linewidth=\LineWidth,linecolor=color320.0012]
(50.000000,0.272677)(100.000000,0.272560)(150.000000,0.375283)(200.000000,0.384765)(250.000000,0.402295)
(300.000000,0.667545)(350.000000,0.681639)(400.000000,0.890399)(450.000000,1.886069)(500.000000,3.377146)
(550.000000,5.069821)(600.000000,6.849207)(650.000000,8.664397)(700.000000,10.517500)(750.000000,12.454021)
(800.000000,14.383826)(850.000000,16.293828)(900.000000,18.164427)(950.000000,19.962195)(1000.000000,22.074333)
(1050.000000,23.769012)(1100.000000,25.893033)(1150.000000,28.000140)

{ \small 
\rput(230,30){%
\psshadowbox[framesep=0pt,linewidth=\AxesLineWidth]{\psframebox*{\begin{tabular}{l}
\Rnode{a1}{\hspace*{0.0ex}} \hspace*{0.4cm} \Rnode{a2}{~~A1} \\
\Rnode{a7}{\hspace*{0.0ex}} \hspace*{0.4cm} \Rnode{a8}{~~B1} \\
\Rnode{a13}{\hspace*{0.0ex}} \hspace*{0.4cm} \Rnode{a14}{~~C1} \\
\end{tabular}}
\ncline[linestyle=solid,linewidth=\LineWidth,linecolor=color246.004]{a1}{a2} \ncput{\psdot[dotstyle=*,dotsize=\MarkerSize,linecolor=color246.004]}
\ncline[linestyle=solid,linewidth=\LineWidth,linecolor=color249.0035]{a7}{a8} \ncput{\psdot[dotstyle=Bo,dotsize=\MarkerSize,linecolor=color249.0035]}
\ncline[linestyle=solid,linewidth=\LineWidth,linecolor=color252.0035]{a13}{a14} \ncput{\psdot[dotstyle=Bsquare,dotsize=\MarkerSize,linecolor=color252.0035]}
}%
}%
} 
{ \small 
\rput(530,30){%
\psshadowbox[framesep=0pt,linewidth=\AxesLineWidth]{\psframebox*{\begin{tabular}{l}

\Rnode{a3}{\hspace*{0.0ex}} \hspace*{0.4cm} \Rnode{a4}{~~A2} \\
\Rnode{a9}{\hspace*{0.0ex}} \hspace*{0.4cm} \Rnode{a10}{~~B2} \\
\Rnode{a15}{\hspace*{0.0ex}} \hspace*{0.4cm} \Rnode{a16}{~~C2} \\
\end{tabular}}
\ncline[linestyle=solid,linewidth=\LineWidth,linecolor=color247.0035]{a3}{a4} \ncput{\psdot[dotstyle=*,dotsize=\MarkerSize,linecolor=color247.0035]}
\ncline[linestyle=solid,linewidth=\LineWidth,linecolor=color250.0035]{a9}{a10} \ncput{\psdot[dotstyle=Bo,dotsize=\MarkerSize,linecolor=color250.0035]}
\ncline[linestyle=solid,linewidth=\LineWidth,linecolor=color253.0035]{a15}{a16} \ncput{\psdot[dotstyle=Bsquare,dotsize=\MarkerSize,linecolor=color253.0035]}
}%
}%
} 
{ \small 
\rput(830,30){%
\psshadowbox[framesep=0pt,linewidth=\AxesLineWidth]{\psframebox*{\begin{tabular}{l}

\Rnode{a5}{\hspace*{0.0ex}} \hspace*{0.4cm} \Rnode{a6}{~~A3} \\
\Rnode{a11}{\hspace*{0.0ex}} \hspace*{0.4cm} \Rnode{a12}{~~B3} \\
\Rnode{a17}{\hspace*{0.0ex}} \hspace*{0.4cm} \Rnode{a18}{~~C3} \\
\end{tabular}}
\ncline[linestyle=solid,linewidth=\LineWidth,linecolor=color248.0035]{a5}{a6} \ncput{\psdot[dotstyle=*,dotsize=\MarkerSize,linecolor=color248.0035]}
\ncline[linestyle=solid,linewidth=\LineWidth,linecolor=color251.0035]{a11}{a12} \ncput{\psdot[dotstyle=Bo,dotsize=\MarkerSize,linecolor=color251.0035]}
\ncline[linestyle=solid,linewidth=\LineWidth,linecolor=color254.0035]{a17}{a18} \ncput{\psdot[dotstyle=Bsquare,dotsize=\MarkerSize,linecolor=color254.0035]}
}%
}%
} 
{ \small 
\rput[tr](1120,8){%
\psshadowbox[framesep=0pt,linewidth=\AxesLineWidth]{\psframebox*{\begin{tabular}{l}
A4, A5, A6\\
B4, B5, B6\\
C4, C5, C6\\
\end{tabular}}
}%
}%
} 
\psellipse[linewidth=\LineWidth,linestyle=dashed,dash=0.06cm 0.06cm](150,11)(30,3)
\psline[linewidth=\LineWidth,linestyle=dashed,dash=0.06cm 0.06cm](150,14)(250,26)
\psellipse[linewidth=\LineWidth,linestyle=dashed,dash=0.06cm 0.06cm](200,7)(30,1)
\psline[linewidth=\LineWidth,linestyle=dashed,dash=0.06cm 0.06cm](200,8)(500,26)
\psellipse[linewidth=\LineWidth,linestyle=dashed,dash=0.06cm 0.06cm](350,7)(30,3)
\psline[linewidth=\LineWidth,linestyle=dashed,dash=0.06cm 0.06cm](350,10)(750,26)
\psellipse[linewidth=\LineWidth,linestyle=dashed,dash=0.06cm 0.06cm](600,7)(30,2)
\psline[linewidth=\LineWidth,linestyle=dashed,dash=0.06cm 0.06cm](600,9)(800,8)
\end{pspicture}%

\caption{The rates allocated $r_{i}^{1}$ from sector 1 in the three cells.}
\label{fig:sim:rates}
\end{figure}
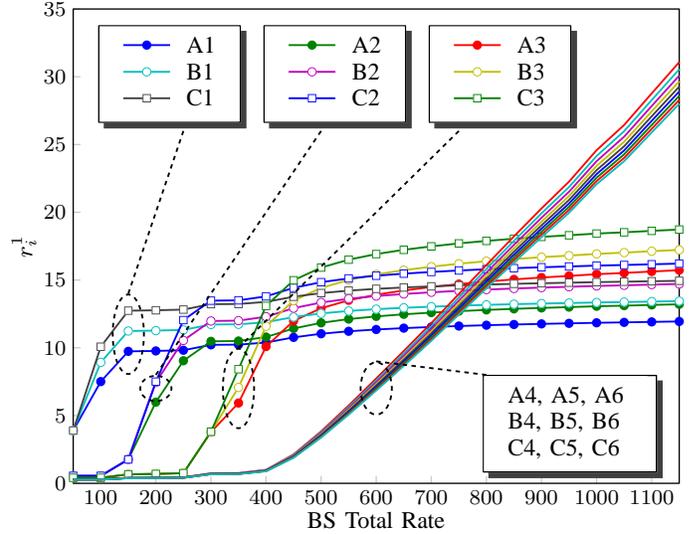
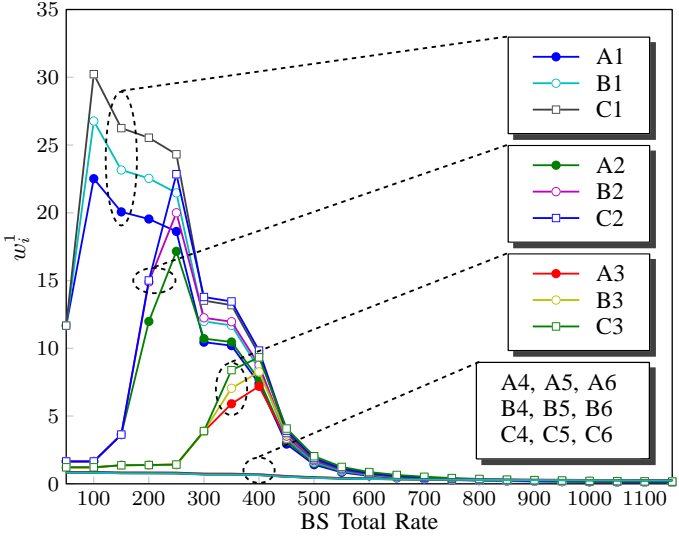
\begin{figure}[]
\centering
%
\psset{xunit=0.000909\plotwidth,yunit=0.022535\plotwidth}%
\begin{pspicture}(-61.520737,-3.888889)(1155.069124,35.818713)%


\psline[linewidth=\AxesLineWidth,linecolor=GridColor](100.000000,0.000000)(100.000000,0.532515)
\psline[linewidth=\AxesLineWidth,linecolor=GridColor](200.000000,0.000000)(200.000000,0.532515)
\psline[linewidth=\AxesLineWidth,linecolor=GridColor](300.000000,0.000000)(300.000000,0.532515)
\psline[linewidth=\AxesLineWidth,linecolor=GridColor](400.000000,0.000000)(400.000000,0.532515)
\psline[linewidth=\AxesLineWidth,linecolor=GridColor](500.000000,0.000000)(500.000000,0.532515)
\psline[linewidth=\AxesLineWidth,linecolor=GridColor](600.000000,0.000000)(600.000000,0.532515)
\psline[linewidth=\AxesLineWidth,linecolor=GridColor](700.000000,0.000000)(700.000000,0.532515)
\psline[linewidth=\AxesLineWidth,linecolor=GridColor](800.000000,0.000000)(800.000000,0.532515)
\psline[linewidth=\AxesLineWidth,linecolor=GridColor](900.000000,0.000000)(900.000000,0.532515)
\psline[linewidth=\AxesLineWidth,linecolor=GridColor](1000.000000,0.000000)(1000.000000,0.532515)
\psline[linewidth=\AxesLineWidth,linecolor=GridColor](1100.000000,0.000000)(1100.000000,0.532515)
\psline[linewidth=\AxesLineWidth,linecolor=GridColor](50.000000,0.000000)(63.200000,0.000000)
\psline[linewidth=\AxesLineWidth,linecolor=GridColor](50.000000,5.000000)(63.200000,5.000000)
\psline[linewidth=\AxesLineWidth,linecolor=GridColor](50.000000,10.000000)(63.200000,10.000000)
\psline[linewidth=\AxesLineWidth,linecolor=GridColor](50.000000,15.000000)(63.200000,15.000000)
\psline[linewidth=\AxesLineWidth,linecolor=GridColor](50.000000,20.000000)(63.200000,20.000000)
\psline[linewidth=\AxesLineWidth,linecolor=GridColor](50.000000,25.000000)(63.200000,25.000000)
\psline[linewidth=\AxesLineWidth,linecolor=GridColor](50.000000,30.000000)(63.200000,30.000000)
\psline[linewidth=\AxesLineWidth,linecolor=GridColor](50.000000,35.000000)(63.200000,35.000000)

{ \footnotesize 
\rput[t](100.000000,-0.532515){$100$}
\rput[t](200.000000,-0.532515){$200$}
\rput[t](300.000000,-0.532515){$300$}
\rput[t](400.000000,-0.532515){$400$}
\rput[t](500.000000,-0.532515){$500$}
\rput[t](600.000000,-0.532515){$600$}
\rput[t](700.000000,-0.532515){$700$}
\rput[t](800.000000,-0.532515){$800$}
\rput[t](900.000000,-0.532515){$900$}
\rput[t](1000.000000,-0.532515){$1000$}
\rput[t](1100.000000,-0.532515){$1100$}
\rput[r](36.800000,0.000000){$0$}
\rput[r](36.800000,5.000000){$5$}
\rput[r](36.800000,10.000000){$10$}
\rput[r](36.800000,15.000000){$15$}
\rput[r](36.800000,20.000000){$20$}
\rput[r](36.800000,25.000000){$25$}
\rput[r](36.800000,30.000000){$30$}
\rput[r](36.800000,35.000000){$35$}
} 

\psframe[linewidth=\AxesLineWidth,dimen=middle](50.000000,0.000000)(1150.000000,35.000000)

{ \small 
\rput[b](600.000000,-3.888889){
\begin{tabular}{c}
BS Total Rate\\
\end{tabular}
}

\rput[t]{90}(-61.520737,17.500000){
\begin{tabular}{c}
$w_i^1$\\
\end{tabular}
}
} 

\newrgbcolor{color95.0018}{0  0  1}
\psline[plotstyle=line,linejoin=1,showpoints=true,dotstyle=*,dotsize=\MarkerSize,linestyle=solid,linewidth=\LineWidth,linecolor=color95.0018]
(50.000000,11.674728)(100.000000,22.517864)(150.000000,20.066236)(200.000000,19.540085)(250.000000,18.623721)
(300.000000,10.451517)(350.000000,10.196638)(400.000000,7.405488)(450.000000,2.931875)(500.000000,1.411161)
(550.000000,0.848031)(600.000000,0.582907)(650.000000,0.435657)(700.000000,0.343143)(750.000000,0.278958)
(800.000000,0.234005)(850.000000,0.201114)(900.000000,0.176320)(950.000000,0.157338)(1000.000000,0.139402)
(1050.000000,0.127560)(1100.000000,0.115135)(1150.000000,0.104864)

\newrgbcolor{color96.0013}{0         0.5           0}
\psline[plotstyle=line,linejoin=1,showpoints=true,dotstyle=*,dotsize=\MarkerSize,linestyle=solid,linewidth=\LineWidth,linecolor=color96.0013]
(50.000000,1.648758)(100.000000,1.648087)(150.000000,3.633686)(200.000000,11.982610)(250.000000,17.155482)
(300.000000,10.714903)(350.000000,10.466398)(400.000000,7.695518)(450.000000,3.110415)(500.000000,1.514041)
(550.000000,0.916065)(600.000000,0.632618)(650.000000,0.474456)(700.000000,0.374738)(750.000000,0.305358)
(800.000000,0.256652)(850.000000,0.220943)(900.000000,0.193980)(950.000000,0.173308)(1000.000000,0.153749)
(1050.000000,0.140820)(1100.000000,0.127239)(1150.000000,0.116000)

\newrgbcolor{color97.0013}{1  0  0}
\psline[plotstyle=line,linejoin=1,showpoints=true,dotstyle=*,dotsize=\MarkerSize,linestyle=solid,linewidth=\LineWidth,linecolor=color97.0013]
(50.000000,1.216559)(100.000000,1.216418)(150.000000,1.368550)(200.000000,1.386185)(250.000000,1.420902)
(300.000000,3.863580)(350.000000,5.904324)(400.000000,7.193308)(450.000000,3.262690)(500.000000,1.651882)
(550.000000,1.020784)(600.000000,0.714697)(650.000000,0.541365)(700.000000,0.430910)(750.000000,0.353411)
(800.000000,0.298627)(850.000000,0.258233)(900.000000,0.227587)(950.000000,0.203996)(1000.000000,0.181590)
(1050.000000,0.166728)(1100.000000,0.151072)(1150.000000,0.138076)

\newrgbcolor{color98.0013}{0        0.75        0.75}
\psline[plotstyle=line,linejoin=1,showpoints=true,dotstyle=Bo,dotsize=\MarkerSize,linestyle=solid,linewidth=\LineWidth,linecolor=color98.0013]
(50.000000,11.674728)(100.000000,26.772445)(150.000000,23.157098)(200.000000,22.540458)(250.000000,21.468648)
(300.000000,11.985531)(350.000000,11.691335)(400.000000,8.474760)(450.000000,3.340284)(500.000000,1.602942)
(550.000000,0.961419)(600.000000,0.659943)(650.000000,0.492722)(700.000000,0.387767)(750.000000,0.315011)
(800.000000,0.264091)(850.000000,0.226855)(900.000000,0.198801)(950.000000,0.177332)(1000.000000,0.157055)
(1050.000000,0.143671)(1100.000000,0.129634)(1150.000000,0.118033)

\newrgbcolor{color103.0027}{0.75           0        0.75}
\psline[plotstyle=line,linejoin=1,showpoints=true,dotstyle=Bo,dotsize=\MarkerSize,linestyle=solid,linewidth=\LineWidth,linecolor=color103.0027]
(50.000000,1.648758)(100.000000,1.648087)(150.000000,3.633687)(200.000000,14.892706)(250.000000,20.000409)
(300.000000,12.248917)(350.000000,11.961095)(400.000000,8.764790)(450.000000,3.518824)(500.000000,1.705822)
(550.000000,1.029453)(600.000000,0.709654)(650.000000,0.531521)(700.000000,0.419361)(750.000000,0.341411)
(800.000000,0.286737)(850.000000,0.246684)(900.000000,0.216461)(950.000000,0.193302)(1000.000000,0.171402)
(1050.000000,0.156931)(1100.000000,0.141738)(1150.000000,0.129169)

\newrgbcolor{color104.0027}{0.75        0.75           0}
\psline[plotstyle=line,linejoin=1,showpoints=true,dotstyle=Bo,dotsize=\MarkerSize,linestyle=solid,linewidth=\LineWidth,linecolor=color104.0027]
(50.000000,1.216568)(100.000000,1.216428)(150.000000,1.368573)(200.000000,1.386211)(250.000000,1.420933)
(300.000000,3.887913)(350.000000,7.049349)(400.000000,8.262468)(450.000000,3.671093)(500.000000,1.843661)
(550.000000,1.134171)(600.000000,0.791732)(650.000000,0.598430)(700.000000,0.475534)(750.000000,0.389463)
(800.000000,0.328712)(850.000000,0.283974)(900.000000,0.250068)(950.000000,0.223989)(1000.000000,0.199242)
(1050.000000,0.182839)(1100.000000,0.165571)(1150.000000,0.151246)

\newrgbcolor{color105.0027}{0.25        0.25        0.25}
\psline[plotstyle=line,linejoin=1,showpoints=true,dotstyle=Bsquare,dotsize=\MarkerSize,linestyle=solid,linewidth=\LineWidth,linecolor=color105.0027]
(50.000000,11.674728)(100.000000,30.229803)(150.000000,26.247960)(200.000000,25.540831)(250.000000,24.313576)
(300.000000,13.519545)(350.000000,13.186031)(400.000000,9.544031)(450.000000,3.748694)(500.000000,1.794723)
(550.000000,1.074808)(600.000000,0.736978)(650.000000,0.549787)(700.000000,0.432391)(750.000000,0.351064)
(800.000000,0.294176)(850.000000,0.252596)(900.000000,0.221281)(950.000000,0.197326)(1000.000000,0.174707)
(1050.000000,0.159783)(1100.000000,0.144133)(1150.000000,0.131203)

\newrgbcolor{color106.0027}{0  0  1}
\psline[plotstyle=line,linejoin=1,showpoints=true,dotstyle=Bsquare,dotsize=\MarkerSize,linestyle=solid,linewidth=\LineWidth,linecolor=color106.0027]
(50.000000,1.648758)(100.000000,1.648087)(150.000000,3.633687)(200.000000,15.005279)(250.000000,22.845336)
(300.000000,13.782931)(350.000000,13.455791)(400.000000,9.834061)(450.000000,3.927234)(500.000000,1.897602)
(550.000000,1.142842)(600.000000,0.786689)(650.000000,0.588586)(700.000000,0.463985)(750.000000,0.377464)
(800.000000,0.316823)(850.000000,0.272425)(900.000000,0.238942)(950.000000,0.213296)(1000.000000,0.189054)
(1050.000000,0.173042)(1100.000000,0.156236)(1150.000000,0.142339)

\newrgbcolor{color107.0027}{0         0.5           0}
\psline[plotstyle=line,linejoin=1,showpoints=true,dotstyle=Bsquare,dotsize=\MarkerSize,linestyle=solid,linewidth=\LineWidth,linecolor=color107.0027]
(50.000000,1.216571)(100.000000,1.216430)(150.000000,1.368579)(200.000000,1.386216)(250.000000,1.420940)
(300.000000,3.893593)(350.000000,8.389673)(400.000000,9.331714)(450.000000,4.079501)(500.000000,2.035441)
(550.000000,1.247559)(600.000000,0.868767)(650.000000,0.655495)(700.000000,0.520158)(750.000000,0.425516)
(800.000000,0.358798)(850.000000,0.309716)(900.000000,0.272549)(950.000000,0.243983)(1000.000000,0.216894)
(1050.000000,0.198951)(1100.000000,0.180070)(1150.000000,0.164416)

\newrgbcolor{color108.0027}{1  0  0}
\psline[plotstyle=line,linejoin=1,linestyle=solid,linewidth=\LineWidth,linecolor=color108.0027]
(50.000000,0.873083)(100.000000,0.873127)(150.000000,0.836215)(200.000000,0.833055)(250.000000,0.827315)
(300.000000,0.754046)(350.000000,0.750751)(400.000000,0.707275)(450.000000,0.580558)(500.000000,0.489509)
(550.000000,0.433863)(600.000000,0.397301)(650.000000,0.371429)(700.000000,0.351797)(750.000000,0.335864)
(800.000000,0.323108)(850.000000,0.312649)(900.000000,0.303953)(1000.000000,0.289277)(1050.000000,0.284003)
(1100.000000,0.278093)(1150.000000,0.272865)

\newrgbcolor{color109.0027}{0        0.75        0.75}
\psline[plotstyle=line,linejoin=1,linestyle=solid,linewidth=\LineWidth,linecolor=color109.0027]
(50.000000,0.865044)(100.000000,0.865091)(150.000000,0.826868)(200.000000,0.823612)(250.000000,0.817705)
(300.000000,0.742953)(350.000000,0.739616)(400.000000,0.695773)(450.000000,0.569517)(500.000000,0.479830)
(550.000000,0.425310)(600.000000,0.389577)(650.000000,0.364324)(700.000000,0.345177)(750.000000,0.329645)
(800.000000,0.317213)(850.000000,0.307022)(900.000000,0.298549)(1000.000000,0.284251)(1050.000000,0.279113)
(1100.000000,0.273355)(1150.000000,0.268262)

\newrgbcolor{color110.0023}{0.75           0        0.75}
\psline[plotstyle=line,linejoin=1,linestyle=solid,linewidth=\LineWidth,linecolor=color110.0023]
(50.000000,0.857369)(100.000000,0.857418)(150.000000,0.818026)(200.000000,0.814687)(250.000000,0.808633)
(300.000000,0.732632)(350.000000,0.729263)(400.000000,0.685151)(450.000000,0.559472)(500.000000,0.471087)
(550.000000,0.417608)(600.000000,0.382632)(650.000000,0.357943)(700.000000,0.339234)(750.000000,0.324063)
(800.000000,0.311923)(850.000000,0.301973)(900.000000,0.293700)(1000.000000,0.279742)(1050.000000,0.274726)
(1100.000000,0.269104)(1150.000000,0.264132)

\newrgbcolor{color111.0023}{0.75        0.75           0}
\psline[plotstyle=line,linejoin=1,linestyle=solid,linewidth=\LineWidth,linecolor=color111.0023]
(50.000000,0.850028)(100.000000,0.850078)(150.000000,0.809641)(200.000000,0.806228)(250.000000,0.800047)
(300.000000,0.722992)(350.000000,0.719597)(400.000000,0.675296)(450.000000,0.550275)(500.000000,0.463131)
(550.000000,0.410619)(600.000000,0.376339)(650.000000,0.352164)(700.000000,0.333854)(750.000000,0.319012)
(800.000000,0.307137)(850.000000,0.297405)(900.000000,0.289314)(1000.000000,0.275662)(1050.000000,0.270757)
(1100.000000,0.265259)(1150.000000,0.260395)

\newrgbcolor{color112.0023}{0.25        0.25        0.25}
\psline[plotstyle=line,linejoin=1,linestyle=solid,linewidth=\LineWidth,linecolor=color112.0023]
(50.000000,0.842996)(100.000000,0.843047)(150.000000,0.801673)(200.000000,0.798195)(250.000000,0.791901)
(300.000000,0.713955)(350.000000,0.710541)(400.000000,0.666113)(450.000000,0.541805)(500.000000,0.455846)
(550.000000,0.404234)(600.000000,0.370597)(650.000000,0.346894)(700.000000,0.328950)(750.000000,0.314408)
(800.000000,0.302775)(850.000000,0.293242)(900.000000,0.285317)(1000.000000,0.271945)(1050.000000,0.267140)
(1100.000000,0.261754)(1150.000000,0.256989)

\newrgbcolor{color113.0023}{0  0  1}
\psline[plotstyle=line,linejoin=1,linestyle=solid,linewidth=\LineWidth,linecolor=color113.0023]
(50.000000,0.836249)(100.000000,0.836301)(150.000000,0.794084)(200.000000,0.790549)(250.000000,0.784156)
(300.000000,0.705457)(350.000000,0.702028)(400.000000,0.657524)(450.000000,0.533968)(500.000000,0.449138)
(550.000000,0.398368)(600.000000,0.365326)(650.000000,0.342059)(700.000000,0.324452)(750.000000,0.310186)
(800.000000,0.298776)(850.000000,0.289425)(900.000000,0.281652)(1000.000000,0.268537)(1050.000000,0.263823)
(1100.000000,0.258540)(1150.000000,0.253866)

\newrgbcolor{color114.0023}{0         0.5           0}
\psline[plotstyle=line,linejoin=1,linestyle=solid,linewidth=\LineWidth,linecolor=color114.0023]
(50.000000,0.829766)(100.000000,0.829820)(150.000000,0.786844)(200.000000,0.783257)(250.000000,0.776777)
(300.000000,0.697443)(350.000000,0.694002)(400.000000,0.649465)(450.000000,0.526685)(500.000000,0.442931)
(550.000000,0.392949)(600.000000,0.360462)(650.000000,0.337600)(700.000000,0.320305)(750.000000,0.306294)
(800.000000,0.295089)(850.000000,0.285907)(900.000000,0.278274)(1000.000000,0.265394)(1050.000000,0.260766)
(1100.000000,0.255577)(1150.000000,0.250987)

\newrgbcolor{color115.0023}{1  0  0}
\psline[plotstyle=line,linejoin=1,linestyle=solid,linewidth=\LineWidth,linecolor=color115.0023]
(50.000000,0.823530)(100.000000,0.823585)(150.000000,0.779923)(200.000000,0.776291)(250.000000,0.769734)
(300.000000,0.689865)(350.000000,0.686417)(400.000000,0.641879)(450.000000,0.519889)(500.000000,0.437162)
(550.000000,0.387922)(600.000000,0.355952)(650.000000,0.333467)(700.000000,0.316462)(750.000000,0.302689)
(800.000000,0.291673)(850.000000,0.282648)(900.000000,0.275145)(1000.000000,0.262484)(1050.000000,0.257933)
(1100.000000,0.252833)(1150.000000,0.248319)

\newrgbcolor{color116.0023}{0        0.75        0.75}
\psline[plotstyle=line,linejoin=1,linestyle=solid,linewidth=\LineWidth,linecolor=color116.0023]
(50.000000,0.817524)(100.000000,0.817580)(150.000000,0.773298)(200.000000,0.769626)(250.000000,0.763000)
(300.000000,0.682682)(350.000000,0.679229)(400.000000,0.634719)(450.000000,0.513526)(500.000000,0.431781)
(550.000000,0.383239)(600.000000,0.351755)(650.000000,0.329622)(700.000000,0.312887)(750.000000,0.299334)
(800.000000,0.288496)(850.000000,0.279616)(900.000000,0.272233)(1000.000000,0.259775)(1050.000000,0.255298)
(1100.000000,0.250279)(1150.000000,0.245837)

{ \small 
\rput[tr](1120,33){%
\psshadowbox[framesep=0pt,linewidth=\AxesLineWidth]{\psframebox*{\begin{tabular}{l}
\Rnode{a1}{\hspace*{0.0ex}} \hspace*{0.4cm} \Rnode{a2}{~~A1} \\
\Rnode{a7}{\hspace*{0.0ex}} \hspace*{0.4cm} \Rnode{a8}{~~B1} \\
\Rnode{a13}{\hspace*{0.0ex}} \hspace*{0.4cm} \Rnode{a14}{~~C1} \\
\end{tabular}}
\ncline[linestyle=solid,linewidth=\LineWidth,linecolor=color95.0018]{a1}{a2} \ncput{\psdot[dotstyle=*,dotsize=\MarkerSize,linecolor=color95.0018]}
\ncline[linestyle=solid,linewidth=\LineWidth,linecolor=color98.0013]{a7}{a8} \ncput{\psdot[dotstyle=Bo,dotsize=\MarkerSize,linecolor=color98.0013]}
\ncline[linestyle=solid,linewidth=\LineWidth,linecolor=color105.0027]{a13}{a14} \ncput{\psdot[dotstyle=Bsquare,dotsize=\MarkerSize,linecolor=color105.0027]}
}%
}%
} 
{ \small 
\rput[tr](1120,25){%
\psshadowbox[framesep=0pt,linewidth=\AxesLineWidth]{\psframebox*{\begin{tabular}{l}
\Rnode{a3}{\hspace*{0.0ex}} \hspace*{0.4cm} \Rnode{a4}{~~A2} \\
\Rnode{a9}{\hspace*{0.0ex}} \hspace*{0.4cm} \Rnode{a10}{~~B2} \\
\Rnode{a15}{\hspace*{0.0ex}} \hspace*{0.4cm} \Rnode{a16}{~~C2} \\
\end{tabular}}
\ncline[linestyle=solid,linewidth=\LineWidth,linecolor=color96.0013]{a3}{a4} \ncput{\psdot[dotstyle=*,dotsize=\MarkerSize,linecolor=color96.0013]}
\ncline[linestyle=solid,linewidth=\LineWidth,linecolor=color103.0027]{a9}{a10} \ncput{\psdot[dotstyle=Bo,dotsize=\MarkerSize,linecolor=color103.0027]}
\ncline[linestyle=solid,linewidth=\LineWidth,linecolor=color106.0027]{a15}{a16} \ncput{\psdot[dotstyle=Bsquare,dotsize=\MarkerSize,linecolor=color106.0027]}
}%
}%
} 
{ \small 
\rput[tr](1120,17){%
\psshadowbox[framesep=0pt,linewidth=\AxesLineWidth]{\psframebox*{\begin{tabular}{l}
\Rnode{a5}{\hspace*{0.0ex}} \hspace*{0.4cm} \Rnode{a6}{~~A3} \\
\Rnode{a11}{\hspace*{0.0ex}} \hspace*{0.4cm} \Rnode{a12}{~~B3} \\
\Rnode{a17}{\hspace*{0.0ex}} \hspace*{0.4cm} \Rnode{a18}{~~C3} \\
\end{tabular}}
\ncline[linestyle=solid,linewidth=\LineWidth,linecolor=color97.0013]{a5}{a6} \ncput{\psdot[dotstyle=*,dotsize=\MarkerSize,linecolor=color97.0013]}
\ncline[linestyle=solid,linewidth=\LineWidth,linecolor=color104.0027]{a11}{a12} \ncput{\psdot[dotstyle=Bo,dotsize=\MarkerSize,linecolor=color104.0027]}
\ncline[linestyle=solid,linewidth=\LineWidth,linecolor=color107.0027]{a17}{a18} \ncput{\psdot[dotstyle=Bsquare,dotsize=\MarkerSize,linecolor=color107.0027]}
}%
}%
} 
{ \small 
\rput[tr](1120,9){%
\psshadowbox[framesep=0pt,linewidth=\AxesLineWidth]{\psframebox*{\begin{tabular}{l}
A4, A5, A6\\
B4, B5, B6\\
C4, C5, C6\\
\end{tabular}}
}%
}%
} 
\psellipse[linewidth=\LineWidth,linestyle=dashed,dash=0.06cm 0.06cm](150,24)(30,5)
\psline[linewidth=\LineWidth,linestyle=dashed,dash=0.06cm 0.06cm](150,29)(850,33)
\psellipse[linewidth=\LineWidth,linestyle=dashed,dash=0.06cm 0.06cm](210,15)(40,1)
\psline[linewidth=\LineWidth,linestyle=dashed,dash=0.06cm 0.06cm](210,16)(850,25)
\psellipse[linewidth=\LineWidth,linestyle=dashed,dash=0.06cm 0.06cm](350,7)(30,2)
\psline[linewidth=\LineWidth,linestyle=dashed,dash=0.06cm 0.06cm](350,9)(850,17)
\psellipse[linewidth=\LineWidth,linestyle=dashed,dash=0.06cm 0.06cm](400,1)(30,1)
\psline[linewidth=\LineWidth,linestyle=dashed,dash=0.06cm 0.06cm](400,2)(800,9)
\end{pspicture}%

\caption{The bids $w_{i}^{1}$ from sector 1 in the three cells.}
\label{fig:sim:bids}
\end{figure}
\section{Simulation Results}\label{sec:sim}

Algorithm (\ref{alg:UE_FK}) , (\ref{alg:BS_FK}) and (\ref{alg:MME_FK}) were applied to various logarithmic and sigmoid utility functions with different parameters in MATLAB. The simulation results showed convergence to the global optimal solution. In this section, we present the simulation results of the users placed in Figure \ref{fig:sim:System_Model}. The utility function parameters for each user is shown in Table \ref{tab:utility}. In the following simulations, we set $\delta =10^{-3}$.

\begin {table}[]
\caption {Users and their utilities} 
\label{tab:utility} 
\begin{center}
\begin{tabular}{| l | l || l | l |}
  \hline
  \multicolumn{4}{|c|}{Sector 1 BS A} \\  \hline
  A1 & Sig $a=3,\:\: b=10.0$ & A4 & Log $k=1.1,\:\:r_{max}=100$  \\ \hline
  A2 & Sig $a=3,\:\: b=10.3$ & A5 & Log $k=1.2,\:\:r_{max}=100$ \\ \hline
  A3 & Sig $a=1,\:\: b=10.6$ & A6 & Log $k=1.3,\:\:r_{max}=100$ \\ \hline
  \multicolumn{4}{|c|}{Sector 2 BS A} \\  \hline
  A7 & Sig $a=3,\:\: b=10$ & A10 & Log $k=1,\:\:r_{max}=100$  \\ \hline
  A8 & Sig $a=3,\:\: b=11$ & A11 & Log $k=2,\:\:r_{max}=100$ \\ \hline
  A9 & Sig $a=1,\:\: b=12$ & A12 & Log $k=3,\:\:r_{max}=100$ \\ \hline  
  \multicolumn{4}{|c|}{Sector 3 BS A} \\  \hline
  A13 & Sig $a=3,\:\: b=15.1$ & A16 & Log $k=10,\:\:r_{max}=100$  \\ \hline
  A14 & Sig $a=3,\:\: b=15.3$ & A17 & Log $k=11,\:\:r_{max}=100$ \\ \hline
  A15 & Sig $a=3,\:\: b=15.5$ & A18 & Log $k=12,\:\:r_{max}=100$ \\ \hline
  
  \multicolumn{4}{|c|}{Sector 1 BS B} \\  \hline
  B1 & Sig $a=3,\:\: b=10.9$ & B4 & Log $k=1.4,\:\:r_{max}=100$  \\ \hline
  B2 & Sig $a=3,\:\: b=11.2$ & B5 & Log $k=1.5,\:\:r_{max}=100$ \\ \hline
  B3 & Sig $a=1,\:\: b=11.5$ & B6 & Log $k=1.6,\:\:r_{max}=100$ \\ \hline
  \multicolumn{4}{|c|}{Sector 2 BS B} \\  \hline
  B7 & Sig $a=3,\:\: b=13$ & B10 & Log $k=4,\:\:r_{max}=100$  \\ \hline
  B8 & Sig $a=3,\:\: b=14$ & B11 & Log $k=5,\:\:r_{max}=100$ \\ \hline
  B9 & Sig $a=1,\:\: b=15$ & B12 & Log $k=6,\:\:r_{max}=100$ \\ \hline  
  \multicolumn{4}{|c|}{Sector 3 BS B} \\  \hline
  B13 & Sig $a=3,\:\: b=15.7$ & B16 & Log $k=13,\:\:r_{max}=100$  \\ \hline
  B14 & Sig $a=3,\:\: b=15.9$ & B17 & Log $k=14,\:\:r_{max}=100$ \\ \hline
  B15 & Sig $a=3,\:\: b=17.3$ & B18 & Log $k=15,\:\:r_{max}=100$ \\ \hline
  
  \multicolumn{4}{|c|}{Sector 1 BS C} \\  \hline
  C1 & Sig $a=3,\:\: b=11.8$ & C4 & Log $k=1.7,\:\:r_{max}=100$  \\ \hline
  C2 & Sig $a=3,\:\: b=12.1$ & C5 & Log $k=1.8,\:\:r_{max}=100$ \\ \hline
  C3 & Sig $a=1,\:\: b=12.4$ & C6 & Log $k=1.9,\:\:r_{max}=100$ \\ \hline
  \multicolumn{4}{|c|}{Sector 2 BS C} \\  \hline
  C7 & Sig $a=3,\:\: b=16$ & C10 & Log $k=7,\:\:r_{max}=100$  \\ \hline
  C8 & Sig $a=3,\:\: b=17$ & C11 & Log $k=8,\:\:r_{max}=100$ \\ \hline
  C9 & Sig $a=1,\:\: b=18$ & C12 & Log $k=9,\:\:r_{max}=100$ \\ \hline  
  \multicolumn{4}{|c|}{Sector 3 BS C} \\  \hline
  C13 & Sig $a=3,\:\: b=17.5$ & C16 & Log $k=16,\:\:r_{max}=100$  \\ \hline
  C14 & Sig $a=3,\:\: b=17.7$ & C17 & Log $k=17,\:\:r_{max}=100$ \\ \hline
  C15 & Sig $a=3,\:\: b=17.9$ & C18 & Log $k=18,\:\:r_{max}=100$ \\ \hline  
\end{tabular}
\end{center}
\end {table}
\subsection{Rates for $50\le R\le1150$}
In Figure \ref{fig:sim:rates}, we show the optimal rates of users in the $1^{st}$ sector versus BS rate $R$. The optimal resource allocation is content-aware. The users with real-time application (i.e. sigmoid utilities) are allocated resources first. In real-time applications allocation, the user with the steepest utility function (largest $a$) is allocated first as shown in Figure \ref{fig:sim:rates}.

\subsection{Bids for $50\le R\le1150$}
In Figure \ref{fig:sim:bids}, we show the optimal bids of users in the $1^{st}$ sector versus BS rate $R$. The users' bids per resource increases as the available resources for allocation in sectors are more scarce, i.e. small values of $R$, as the number of users is fixed. Therefore, we have a traffic-dependent pricing. Provided this traffic-dependent pricing, the network providers can flatten the traffic specially during peak hours by setting traffic-dependent bandwidth resource price, which gives an incentive for users to use the network during less traffic hours.

\section{Conclusion}\label{sec:conclude}
In this paper, we formulated a novel rate allocation optimization problem with frequency reuse. We considered mobile users running two different types of applications, i.e. real-time and delay-tolerant applications, with utility proportional fairness allocation policy in cellular networks. We proved that the global optimal solution exists and is tractable for mobile stations running delay-tolerant and real-time applications. We presented a distributed algorithm, running on UE, BS and MME, for allocating resources to mobile users in different cells and sectors optimally. Our algorithm ensures fairness in utility percentage achieved by the allocated resources for all users. Therefore, the algorithm gives priority to the users with adaptive real-time applications with providing a minimum QoS for all service subscribers. We showed through simulations that our algorithm converges to the optimal rates.

\bibliographystyle{ieeetr}
\bibliography{pubs}

\begin{thebibliography}{10}

\bibitem{DBLP:conf/pimrc/TychogiorgosGL11}
G.~Tychogiorgos, A.~Gkelias, and K.~K. Leung, ``Towards a fair non-convex
  resource allocation in wireless networks,'' in {\em PIMRC}, pp.~36--40, 2011.

\bibitem{DBLP:conf/globecom/TychogiorgosGL11}
G.~Tychogiorgos, A.~Gkelias, and K.~K. Leung, ``A new distributed optimization
  framework for hybrid ad-hoc networks,'' in {\em GLOBECOM Workshops},
  pp.~293--297, 2011.

\bibitem{DL_PowerAllocation}
J.-W. Lee, R.~R. Mazumdar, and N.~B. Shroff, ``Downlink power allocation for
  multi-class wireless systems,'' {\em IEEE/ACM Trans. Netw.}, vol.~13,
  pp.~854--867, Aug. 2005.

\bibitem{RebeccaThesis}
R.~L. Kurrle, ``{Resource Allocation for Smart Phones in 4G LTE Advanced
  Carrier Aggregation},'' {Master Thesis}, {Virginia Tech}, Nov. 2012.

\bibitem{multi_cell_RA}
E.~Björnson and E.~Jorswieck, ``Optimal resource allocation in coordinated
  multi-cell systems,'' {\em Foundations and Trends® in Communications and
  Information Theory}, vol.~9, no.~2–3, pp.~113--381, 2012.

\bibitem{RA_interference}
M.~Hong and Z.~Luo, ``Signal processing and optimal resource allocation for the
  interference channel,'' {\em CoRR}, vol.~abs/1206.5144, 2012.

\bibitem{opt_net_ref}
W.~Utschick and J.~Brehmer, ``Monotonic optimization framework for coordinated
  beamforming in multicell networks,'' {\em Signal Processing, IEEE
  Transactions on}, vol.~60, pp.~1899--1909, April 2012.

\bibitem{Ahmed_Utility1}
A.~Abdel-Hadi and C.~Clancy, ``{A Utility Proportional Fairness Approach for
  Resource Allocation in 4G-LTE},'' in {\em ICNC Workshop CNC}, 2014.

\bibitem{Ahmed_Utility2}
A.~Abdel-Hadi and C.~Clancy, ``{A Robust Optimal Rate Allocation Algorithm and
  Pricing Policy for Hybrid Traffic in 4G-LTE},'' in {\em PIMRC}, 2013.

\bibitem{Ahmed_Utility3}
A.~Abdel-Hadi, C.~Clancy, and J.~Mitola, ``{A Resource Allocation Algorithm for
  Multi-Application Users in 4G-LTE},'' in {\em MobiCom Workshop}, 2013.

\bibitem{GAC_MILCOM}
M.~Ghorbanzadeh, A.~Abdelhadi, and C.~Clancy, ``A utility proportional fairness
  resource allocation in spectrally radar-coexistent cellular networks,'' in
  {\em Military Communications Conference (MILCOM), 2014 IEEE}, pp.~1498--1503,
  Oct 2014.

\bibitem{UtilityFairness}
G.~Tychogiorgos, A.~Gkelias, and K.~K. Leung, ``Utility-proportional fairness
  in wireless networks.,'' in {\em PIMRC}, pp.~839--844, IEEE, 2012.

\bibitem{SAC_CA_user_discrimination}
H.~Shajaiah, A.~Abdelhadi, and C.~Clancy, ``Multi-application resource
  allocation with users discrimination in cellular networks,'' {\em CoRR},
  vol.~abs/1406.1818, 2014.

\bibitem{Haya_joint}
H.~Shajaiah, A.~Abdelhadi, and T.~C. Clancy, ``Robust resource allocation with
  joint carrier aggregation for multi-carrier cellular networks,'' {\em CoRR},
  vol.~abs/1503.08994, 2015.

\end{thebibliography}
\end{document}